\documentclass{aa}
\usepackage{graphicx}
\usepackage{txfonts}
\usepackage{natbib}
\bibpunct{(}{)}{;}{a}{}{,} % to follow the A&A style
\begin{document} 
   \title{\emph{Herschel} survey of brown dwarf disks in $\rho$ Ophiuchi
        \thanks{\emph{Herschel} is an ESA space observatory with science instruments provided by European-led Principal Investigator consortia and with important participation from NASA.}}
   \author{C. Alves de Oliveira\inst{1}
          \and
          P. \'Abrah\'am\inst{2}
          \and
         G. Marton\inst{2}
          \and
          C. Pinte \inst{3}
          \and
          Cs. Kiss\inst{2}
          \and
          M. Kun \inst{2}
          \and
          \'A. K\'osp\'al \inst{4}
          \and
          P. Andr\'e\inst{5}
          \and
          V. K\"onyves\inst{5,6}
          }

   \institute{European Space Astronomy Centre (ESA/ESAC), P.O. Box, 78, 28691 Villanueva de la Ca\~{n}ada, Madrid, Spain \\
   \email{calves@sciops.esa.int}
    \and 
    Konkoly Observatory, Research Centre for Astronomy and Earth Sciences, Hungarian Academy of Sciences, PO Box 67, 1525 Budapest, Hungary
         \and
UJF-Grenoble 1/CNRS-INSU, Institut de Plan\'etologie et d'Astrophysique de Grenoble (IPAG) UMR5274, Grenoble,38041, France
         \and  
           European Space Research and Technology Centre (ESA/ESTEC), P.O. Box 299, 2200 AG Noordwijk, The Netherlands
	\and
	Laboratoire AIM, CEA/DSM-CNRS-Universit\'e Paris Diderot, IRFU/Service d'Astrophysique, C.E. Saclay, Orme des Merisiers, 91191 Gif-sur-Yvette, France
	\and
       IAS, CNRS (UMR 8617), Universit\'{e} Paris-Sud 11, B\^{a}timent 121, 91400 Orsay, France}
   \date{Received 30 July 2013  / Accepted 26 September 2013 }

  \abstract
  {Young brown dwarfs are known to possess circumstellar disks, a characteristic that is fundamental to the understanding of their formation process, and raises the possibility that these objects harbour planets.} 
  {We want to characterise the so far scarcely studied far-IR emission of disks around the young brown dwarf population of the $\rho$~Ophiuchi cluster in \object{LDN~1688}.} 
  {Recent observations of the $\rho$ Ophiuchi cluster with the \emph{Herschel} Space Observatory allow us to probe the spectral energy distribution (SED) of the brown dwarf population in the far-IR, where the disk emission peaks. We performed aperture photometry at 70, 100, and 160~$\mu$m, and constructed SEDs for all previously known brown dwarfs detected. These were complemented with ancillary photometry at shorter wavelengths. We compared the observed SEDs to a grid of synthetic disks produced with the radiative transfer code MCFOST, and used the relative figure of merit estimated from the Bayesian inference of each disk parameter to analyse the structural properties.} 
  {We detected 12 Class~II brown dwarfs with \emph{Herschel}, which corresponds to one-third of all currently known brown dwarf members of $\rho$~Ophiuchi. We do not detect any of the known Class~III brown dwarfs. Comparison to models reveals that the disks are best described by an inner radius between 0.01 and 0.07~AU, and a flared disk geometry with a flaring index between 1.05 and 1.2. Furthermore, we can exclude values of the disk scale-height lower than 10~AU (measured at a fiducial radius of 100~AU). We combined the \emph{Herschel} data with recent ALMA observations of the brown dwarf GY92~204 (ISO$-$Oph~102), and by comparing its SED to the same grid of disk models, we derived an inner disk radius of 0.035~AU, a scale height of 15~AU with a flaring index of $\beta\sim$1.15, an exponent for dust settling of $-$1.5, and a disk mass of 0.001~$M_{\sun}$. This corresponds to a disk-to-central object mass ratio of $\sim$1$\%$.} 
  {The structural parameters constrained by the extended SED coverage (inner radius and flaring index) show a narrow distribution for the 11 young brown dwarfs detected in $\rho$~Ophiuchi, suggesting that these objects share the same disk evolution and, perhaps, formation.}
  {}
\renewcommand{\tabcolsep}{3pt}
\maketitle
\begin{table*}[!ht]
\caption{\emph{Herschel}/PACS fluxes for the detected brown dwarfs in $\rho$~Ophiuchi (upper limits for non-detections are listed in Table~\ref{table2}).}             
\label{table1}      
\centering          
\begin{tabular}{l c c c c c c c c}    
\hline\hline       
Identifier & RA & Dec & SpT  &  A$_{\emph{V}}$ & Ref. & 70~$\mu$m & 100~$\mu$m & 160~$\mu$m \\ 
  	      &       &         &          &    (mag)         	 & 						& (mJy) & (mJy) & (mJy)   \\ 
\hline                    
\object{GY92~154}	        & 16:26:54.79 & $-$24:27:02.1 & M6    			& 20.1 & 1,4,8	  	&  87.5$\pm$10.3     & 172.8$\pm$32.8   & 399.6$\pm$246.4 \\   
\object{GY92~171}	        & 16:26:58.41 & $-$24:21:30.0 & M6    			& 6.6  & 1,4,8	  	&  41.2$\pm$10.5     	 & 53.6$\pm$34.8     & 	  $<$189.0 \\   
\object{GY92~204}\tablefootmark{a}	        & 16:27:06.60 & $-$24:41:48.8 & M6(M5.5)     	& 0.5  & 1,4,5,6 	&  80.1$\pm$5.6      	 & 48.4$\pm$19.1    & 	  $<$115.2	\\   
\object{ISO-Oph~160}        	& 16:27:37.42 & $-$24:17:54.9 & M6    			& 6.0  & 4,5	  	&  38.8$\pm$7.6      	 & 61.9$\pm$20.5     & 	  $<$93.8 	\\   
\object{GY92~344}	        & 16:27:45.78 & $-$24:44:53.6 & M6    			& 16.2 & 1,4,8	  	&  727.3$\pm$4.4     	 & 1338.8$\pm$12.8   & 2675.6$\pm$87.4	\\   
\object{GY92~371}        	& 16:27:49.77 & $-$24:25:22.2 & M6    			& 5.4  & 1,4,8	  	&  42.5$\pm$6.3      	 & 68.4$\pm$19.8    & 	  $<$139.7		\\   
\object{GY92~397}	        & 16:27:55.24 & $-$24:28:39.7 & M6    			& 5.0  & 1,4,8	  	&  64.5$\pm$6.1      	 & 74.7$\pm$18.4     & 	  $<$103.4		\\   
\object{ISO-Oph~193}       	& 16:28:12.72 & $-$24:11:35.6 & M6    			& 7.5  & 4,5	  	&  43.2$\pm$4.6      	 & 38.6$\pm$11.8     & 	  63.2$\pm$58.8 	\\   
\object{CFHTWIR$-$Oph~66}   	& 16:27:14.34 & $-$24:31:32.0 & M7.75 		& 15.1 & 7,9 	&  46.2$\pm$9.6     	&  51.6$\pm$28.2     &	  $<$201.0	       \\  
\object{GY92~3}	       		& 16:26:21.90 & $-$24:44:39.8 & M8(M7.5)  		& 0.0  & 1,4,5,6  	&  45.6$\pm$6.1      	&  $<$27.2    &	  $<$65.7	       \\  
\object{GY92~264} 	       	& 16:27:26.58 & $-$24:25:54.4 & M8    			& 0.0  & 1,6	     	&  26.1$\pm$8.6     	&  64.1$\pm$28.4    &	  $<$158.3	       \\  
\object{GY92~310}	       	& 16:27:38.63 & $-$24:38:39.2 & M8.5(M7,M6) 	& 5.7  & 1,2,3,4,5	&  93.8$\pm$5.5     	&  80.8$\pm$15.8    &	  $<$119.4	       \\  
\hline                  
\end{tabular}
\tablefoot{\tablefoottext{a}{GY92~204 corresponds to \object{ISO$-$Oph~102} observed with ALMA by \citet{Ricci2012}.}}
\tablebib{(1)~\citet{Greene1992}; (2)~\citet{Wilking1999}; (3) \citet{Luhman1999}; (4)~\citet{Bontemps2001}; (5)~\citet{Natta2002}; (6)~\citet{Wilking2005}; (7)~ \citet{AlvesdeOliveira2010}; (8)~\citet{McClure2010}; (9)~\citet{AlvesdeOliveira2012}.}
\end{table*}

\section{Introduction}
Since the prediction of the existence of brown dwarfs \citep{Hayashi1963,Kumar1963} and the first detections \citep{Rebolo1995,Nakajima1995,Oppenheimer1995}, hundreds of these objects have been observed and characterised in a plethora of environments, from star forming regions to the field. The formation mechanism of brown dwarfs remains a subject of debate, with a range of theories being proposed where they form either as an extension of the star formation process \citep{Padoan2002,Hennebelle2008}, or by separate mechanisms such as gravitational instabilities in disks \citep{Stamatellos2009,Basu2012}, premature ejection from prestellar cores \citep{Reipurth2001}, or photo-erosion of cores \citep{Kroupa2003}. 

The observational properties of brown dwarfs are fundamental to the understanding of their formation. For example, the recent discovery of a pre-brown dwarf by \citet{Andre2012} using millimetre interferometric observations has lent additional evidence that brown dwarfs are capable of forming in the same way as solar-mass stars. Furthermore, observations of brown dwarfs at their early evolutionary stages suggest that they undergo a T~Tauri phase analogous to that of hydrogen-burning stars. For example, brown dwarfs are surrounded by circumstellar disks \citep[e.g.,][]{Muench2001}, undergo accretion \citep[e.g.,][]{Mohanty2005}, drive molecular outflows \citep[e.g.,][]{Whelan2009}, and are magnetically active showing evidence of rotational modulation of spots on their surface \citep[e.g.,][]{Rodriguez2009}.  It is therefore pertinent to ask whether brown dwarfs may also form and harbour planetary systems. Searches of planets around cool dwarfs in the field have failed to find any extrasolar planets orbiting these objects \citep[e.g.,][]{Muirhead2012,Kubas2012}. However, \citet{Han2013} have recently announced the discovery of a planetary-mass object orbiting a field brown dwarf using gravitation microlensing. By studying the disks of the young analogues of these objects in the earlier stages of formation, we can assess whether the conditions to form planets are present. 

In this paper, we aim at studying the disks of the brown dwarf population of the $\rho$~Ophiuchi cluster (\object{LDN~1688}). At 1~Myr, the age usually taken for $\rho$~Oph, the hydrogen-burning mass limit corresponds to a spectral type of $\sim$M6.25 \citep{Luhman2007,Baraffe1998,Chabrier2000}. We focus our work on the most updated list of spectroscopically confirmed members with spectral type later than $\sim$M6 compiled in \citet{AlvesdeOliveira2012}, adding to this list one brown dwarf found by \citet{Muzic2012}. There are 43 spectroscopically confirmed brown dwarf members of Ophiuchus. The existence of disks around several of these objects has been previously established from the detection of mid-IR excess \citep[e.g.,][]{Bontemps2001,Natta2002,Testi2002,Barsony2005,McClure2010,AlvesdeOliveira2012}. The T~Tauri disk population of this cluster has been studied by previous missions extending to the far-infrared regime \citep[e.g., IRAS, ISO, and \emph{Spitzer},][]{Young1986,Bontemps2001,Wilking2001,Padgett2008}. However, even with the increased sensitivity of \emph{Spitzer} over ISO, none of the brown dwarf disks was detected by \citet{Padgett2008} in the analysis of the MIPS 70~$\mu$m data \citep[see also][]{Evans2009}, likely due to limitations imposed by the bright background and poor angular resolution. 

We present new photometric observations at 70, 100, and 160~$\mu$m obtained with the \emph{Herschel} Space Observatory \citep{Pilbratt2010}, extending the SED of the young brown dwarfs to the far-IR regime to study the physical properties of their disks. At the time of the writing, this study presents the first complete survey of brown dwarf disks in a young cluster with the \emph{Herschel} Space Observatory at far-IR wavelengths. Previous comparable works include a survey of brown dwarf and low mass star disks sampling few objects per cluster across different environments and ages \citep{Harvey2012b,Harvey2012a}. Prior to the launch of \emph{Herschel}, some benchmark studies of disks around brown dwarfs have been conducted in the submillimeter \citep{Scholz2006,Bouy2008,Joergens2012}. More recently, \citet{Ricci2012} have used ALMA to observe a brown dwarf in $\rho$~Oph. In Sect.~\ref{obs} we present the \emph{Herschel} observations and describe the data reduction. In Sect.~\ref{results} we show the brown dwarf detections in our photometric survey, and the modelling of their disks. We discuss the implications of our results in Sect.~\ref{discussion}, presenting the conclusions in Sect.~\ref{conclusions}. 
%__________________________________________________________________

\begin{figure*}[ht]
  \centering
 \includegraphics[width=\linewidth]{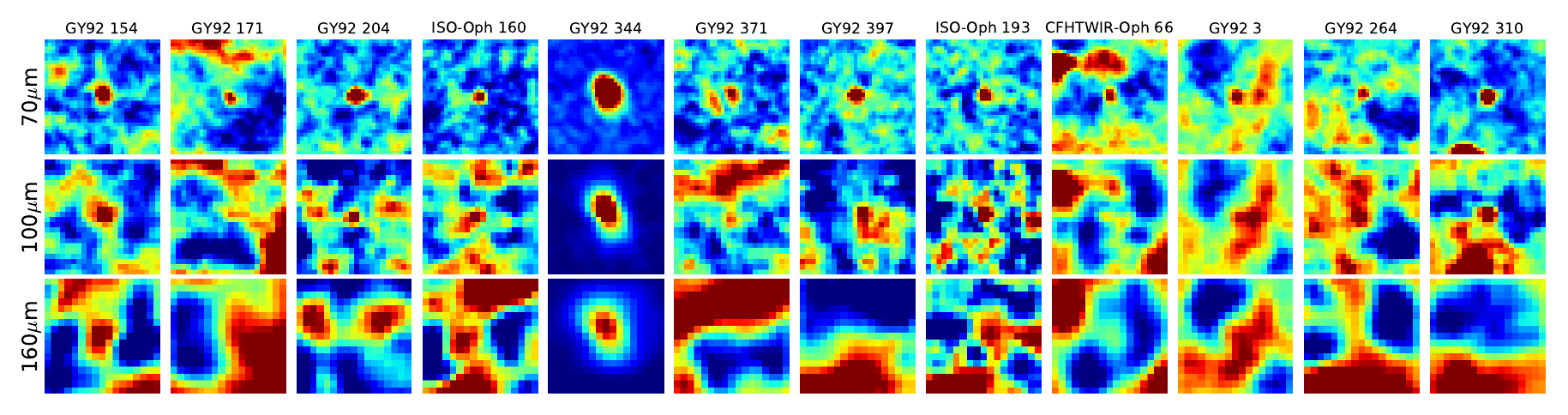}
 \caption{PACS/\emph{Herschel} images of young brown dwarfs in $\rho$~Ophiuchi at 70~$\mu$m \emph{(top row)}, 100~$\mu$m \emph{(middle row)}, and 160~$\mu$m \emph{(bottom row)}. All images are 60$\arcsec$ by 60$\arcsec$ in size centred on the brown dwarf's coordinates (North is up, East is left), are shown in ``DS9 Zscale'' stretch colour, and, for display clarity only, have been smoothed by a 1~$\sigma$ gaussian filter.}
\label{stamps}
\end{figure*}

\section{Observations and data reduction} \label{obs}
The $\rho$~Ophiuchi cluster was observed with the PACS instrument \citep{Poglitsch2010} onboard \emph{Herschel} on February 8, 2012 (program OT1\_pabraham\_3). Observations were made in PACS scan mapping mode (70 and 160~$\mu$m) using a scan and cross-scan speed of 20$\arcsec$$/s$ in order to reach optimum point-source sensitivity and PSF stability, totalling 12.8 hours of observing time. The map covers an area of $\sim$1.6~deg$^{2}$ centred on the Ophiuchus cluster (16$^{h}$27$^{m}$06.30$^{s}$, $-$24$\degr$28$\arcmin$48.5$\arcsec$). We have added to this study data from the \emph{Herschel} Gould Belt survey \citep[HGBS,][]{Andre2010} taken with the PACS scan mapping mode (100 and 160~$\mu$m), with an observing time of 7 hours for a map slightly larger on sky than ours.

The data reduction for the OT1 and HGBS observations was done from ``level 0'' to ``level 2.5'' (final map with the combined scan directions) within the \emph{Herschel} Interactive Processing Environment \citep[HIPE version 8.2,][]{Ott2010}, where an optimised pipeline was created tailored for faint source detection. For the 160~$\mu$m data, we combined the OT1 and HGBS observations to achieve the deepest possible map. In the preprocessing stage, we used scan-speed selection, where the allowed velocities were in the range of 10 to 30$\arcsec$$/s$. Then, a high-pass filter was applied with a filter-width of 15, 16, and 32~$readouts$, and a pixel size of 2, 3, and 4$\arcsec$ was used to project blue, green, and red maps (70, 100, 160~$\mu$m), respectively, using the photProject map making tool. The flux cut level for high-pass-filter masking was set to be 1.5 times the standard deviation of the per-pixel flux values. Glitches were removed with the built-in MMT routine. Although the MMT deglitching method is better suited for faint targets, it may cause flux loss in the case of bright sources, typically for those above 100~mJy. To check whether this phenomenon can modify our values, an additional set of maps was created, now using second level deglitching, instead of MMT. The fluxes obtained by the two methods were compatible within the photometric uncertainties, even for our brightest sources (GY92~154 and GY92~344) and therefore we used the MMT flux values throughout the paper. 

We searched the PACS maps for detections of the 43 known brown dwarfs and the pre-brown dwarf \citep{Andre2012} in the cluster by attempting to fit a 2D Gaussian within $<$4$\arcsec$ from each source's expected coordinates. For sources where this was successful, the point-source flux was obtained by doing aperture photometry centred on the 2D Gaussian peak coordinates. The aperture radii were set to be the average value between the ellipse axes of the PSF full-width half-maximum (5.61, 6.79, and 11.39$\arcsec$, for 70, 100, 160~$\mu$m, respectively). The interpolated encircled energy fraction (EEF) for a given aperture size was calculated by fitting a hexic equation to the EEF values reported in the PACS Point-Source Flux Calibration TN (April 2011). For the sky subtraction, we used two radii of 2 and 4 times the aperture radius. Finally, the colour correction values reported in the PACS Photometer Passbands and Colour Correction Factors for Various Source SEDs (April 2011) were used to correct the flux for all bands (for a blackbody temperature of $\sim$1000~K). We validated this method by calculating the flux of reference stars, for which we find an average difference from the reference values of $\sim$2\%. To estimate the uncertainty of the flux measurements, we populated the images with artificial sources of 1~Jy brightness in a radius of 10$\arcmin$ from each source position. A source mask was first created with \emph{GetSources} \citep{Menshchikov2012}, to ensure that there was no overlap between artificial and real sources. Aperture photometry was carried out in the same way as for the brown dwarfs. This process was repeated 200 times for each brown dwarf location, with the measured fluxes varying around 1~Jy. We take the standard deviation of the 200 artificial source fluxes as a measure of the photometric uncertainty that includes the contribution of the sky background, as well as that of the photometric extraction method. The fluxes of the brown dwarfs detected with PACS are given in Table~\ref{table1}. All detected sources have a 3~$\sigma$ detection in at least one PACS band. In Table~\ref{table2} we give the 1~$\sigma$ upper-limits for the non-detected sources.

The HGBS observations were taken in parallel mode, therefore we could also search the observations taken with the SPIRE instrument \citep{Griffin2010} for counterparts of the Ophiuchus brown dwarfs. Only the source GY92~344 is detected. Using aperture photometry we derived flux densities of 3.90, 4.20, and 2.26~Jy at 250, 350, and 500~$\mu$m (aperture radii of 22, 30 and 42$\arcsec$), respectively. We did not apply an aperture correction to these flux values since there is a strong contribution from scattered light (see Sect.~\ref{sed_results}).
%__________________________________________________ 

\begin{figure*}[ht]
   \centering
 \includegraphics[width=\linewidth]{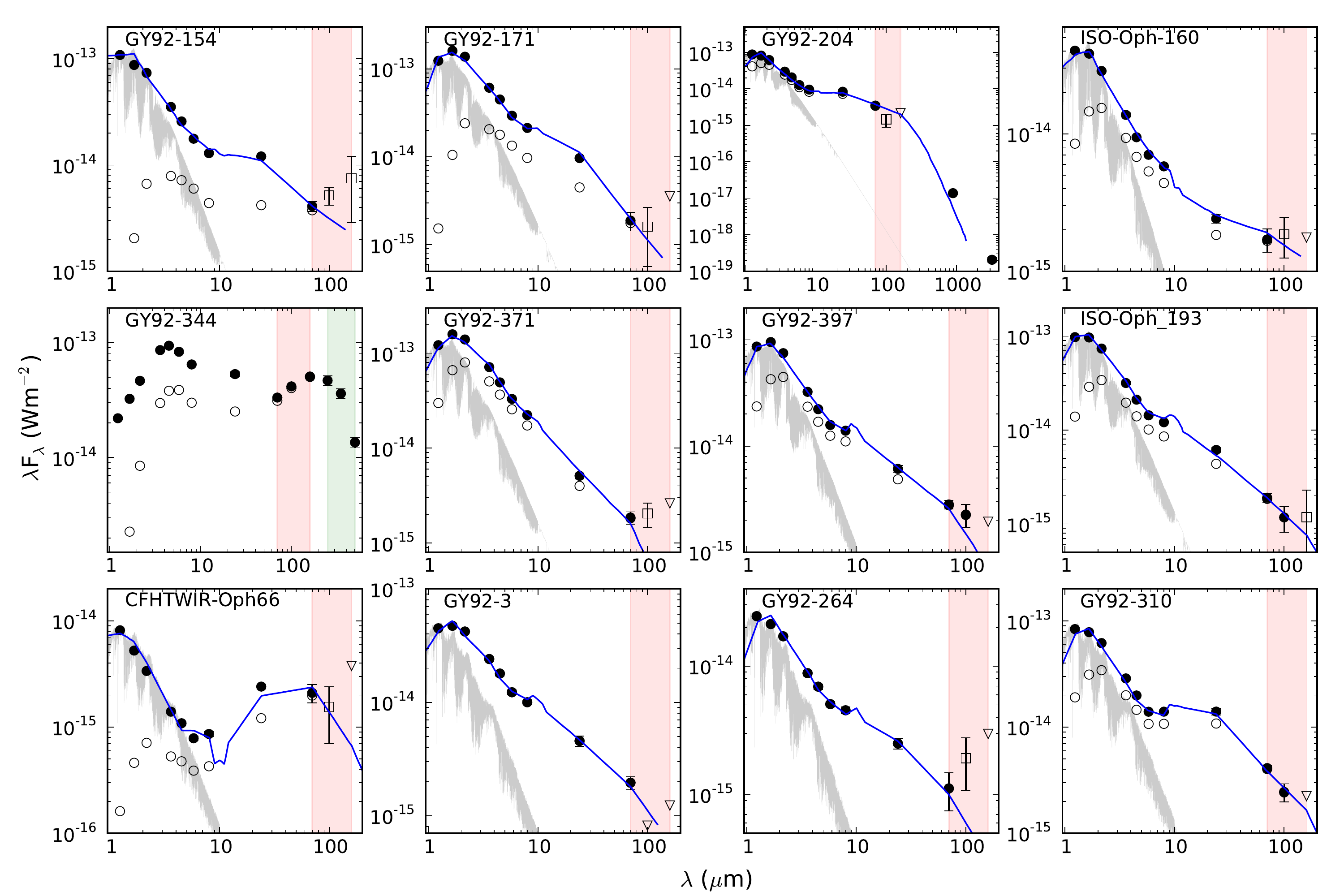}
\caption{SEDs of the young brown dwarfs detected with \emph{Herschel}. All fluxes have been dereddened using the visual extinction values from Table~\ref{table1} and are shown as filled circles for $>$3~$\sigma$ detections, open squares depict marginal detections (1$<$~$\sigma$~$<$3), and open triangles show 1~$\sigma$ upper limits. The original fluxes are marked with open circles. The PACS/\emph{Herschel} wavelength range is highlighted by the red shaded region, and for GY92~344 the SPIRE/\emph{Herschel} wavelength range is marked by the green shaded region. NextGen model photospheres are shown in light grey. The model with the best fit parameters is overplotted in blue.}
\label{sed}
\end{figure*}

\section{Results}  \label{results}

\subsection{PACS/\emph{Herschel} detections}
Previous mid-IR studies of $\rho$~Ophiuchi have classified 31 brown dwarfs as Class~II \citep{Bontemps2001,Natta2002,Testi2002,Barsony2005,McClure2010,AlvesdeOliveira2012}. We detect 38\% of the Class~II brown dwarfs in our deep \emph{Herschel} observations. In particular, we detect 12 sources at 70~$\mu$m with S/N~$>$~3$\sigma$. From these, we detect 11 sources at 100~$\mu$m, but with a lower S/N which results from the shorter exposure time of those observations. Only 3 sources are detected at 160~$\mu$m, where background emission becomes higher. Figure~\ref{stamps} shows the PACS/\emph{Herschel} images of the detected young brown dwarfs in $\rho$~Ophiuchi. 

\citet{Harvey2012a} included in their sample of brown dwarfs one member of Ophiuchus that is also located within our PACS/\emph{Herschel} map (CFHTWIR-Oph~96). We do not detect this object in our deep observations and only derive the 1~$\sigma$ upper-limits of $<$23.5, $<$113.6, and $<$396.3~mJy at 70, 100, and 160~$\mu$m, respectively. These values are in agreement with those measured by \citet{Harvey2012a} from their PACS ``mini-scan-maps'' (F$_{\nu}$ (70~$\mu$m)~$=$~5$\pm$20~mJy and F$_{\nu}$ (160~$\mu$m)~$=$~130$\pm$200~mJy).

The recently discovered pre-brown dwarf core in the Ophiuchus cluster by \citet{Andre2012} using millimetre interferometric observations (\object{Oph~B-11}) is not detected in our PACS maps and we are only able to derive the 1~$\sigma$ upper-limits of $<$15.4, $<$32.0, and $<$63.1~mJy at 70, 100, and 160~$\mu$m, respectively.

\subsection{Spectral energy distribution of detected brown dwarfs} \label{sed_results}
We constructed the SED for each source detected with PACS by adding ancillary data at near and mid-IR wavelengths, from 2MASS \citep{Cutri2003}, WIRCam/CFHT, and \emph{Spitzer} \citep{AlvesdeOliveira2010}. Figure~\ref{sed} shows the final SEDs with the original fluxes plotted as open circles. The black filled circles (for $>$3~$\sigma$ detections) show the fluxes dereddened by the extinction values listed in Table~\ref{table1}, using the extinction law of \citet{Weingartner2001}. The wavelength range of the new PACS/\emph{Herschel} observations is highlighted in red. Several sources are only marginally detected at 100 and 160~$\mu$m (1~$<$~$\sigma$~$<$~3), and these are shown as open squares. The 1~$\sigma$ upper-limits are shown as inverted triangles. 

For two sources (GY92~204 and 344), the SEDs are further complemented at longer wavelengths. The brown dwarf GY92~204 has recently been observed in the millimetre with ALMA \citep{Ricci2012}. They measure a flux density of 4.10$\pm$0.22~mJy at 0.89~mm and 0.22$\pm$0.03~mJy at 3.2 mm. We include those measurements on the SED. The other source, GY92-344 is detected at all SPIRE wavelengths. From near-IR images and the shape of its SED with a significant IR excess, this object resembles a very massive edge-on disk. However, it is most likely an example of a large shadow projected by its circumstellar disk on the surrounding cloud  \citep[see, for example,][]{Hodapp2004,Pontoppidan2005}. In the online Fig.~\ref{shadow}, we show a colour-composite of the \emph{JHK} WIRCam/CFHT near-IR images, where the central object and scattered light are clearly seen. A detailed study of this object is beyond the scope of this paper. 

\begin{table*}
\caption{Range of validity from the Bayesian inference for the disk model parameters (in cols. 3 to 9, the first number corresponds to the model plotted in Fig.~\ref{sed}).}             
\label{table3}      
\centering          
   \tiny
\begin{tabular}{@{}l@{\hspace{0.2cm}}l @{\hspace{0.2cm}}l @{\hspace{0.2cm}}l @{\hspace{0.2cm}}l @{\hspace{0.2cm}}l @{}l @{\hspace{0.2cm}}l @{\hspace{0.2cm}}l @{\hspace{0.2cm}}l@{ }}    
\hline\hline       
Identifier	&	M$_{BD}$\tablefootmark{a}	& Inclination	& Dust mass	& $R_{in}$ & $h_{0}$@100AU\tablefootmark{b}		& $\beta$\tablefootmark{b}	& p\tablefootmark{b} & amax	\\
\hline                  
 	   & ($M_{Jup}$) & (${\degr}$)	& Log$_{10}$($M/M_{\sun}$) &  (AU) & & & & ($\mu$m) \\
\hline
GY92~154		&	75$...$121	&	41~[0:46]		&	$-$4~[$-$6:$-$4]		&	0.0354~[0.0096:0.0679]	&	10~[10:20]		&	1.1~[1.1:1.15]		&	$-$1.5~[$-$1.5:$-$1.5]	&	1000~[1:1000] \\
GY92~171		&	75$...$121	&	32~[0:60]		&	$-$4.5~[$-$7:$-$5]		&	0.0184~[0.0096:0.0354]	&	10~[10:20]		&	1.0~[1.0:1.1]		&	$-$1.5~[$-$1.5:$-$0.5]	&	1000~[1:1000] \\
GY92~204		&	75$...$121	&	63~[0:66]		&	$-$5~[$-$5:$-$5]		&	0.0354~[0.0184:0.0354]	&	15~[10:20]		&	1.15~[1.0:1.15]		&	$-$1.5~[$-$1.5:$-$1.0]	&	1~[1:1] \\
ISO-Oph~160		&	75$...$121	&	70~[0:66]		&	$-$5~[$-$6.5:$-$4]		&	0.005~[0.005:0.0679]	&	15~[10:20]		&	1.05~[1.05:1.15]	&	$-$0.5~[$-$1.5:$-$0.5]	&	1000~[1:1000] \\
GY92~371AB		&	75$...$121	&	70~[26:76]		&	$-$6.5~[$-$7:$-$5]		&	0.0184~[0.0095:0.0354]	&	10~[10:15]		&	1.0~[1.0:1.05]		&	$-$1~[$-$1.5:$-$1.0]	&	1~[1000:1000] \\
GY92~397		&	75$...$121	&	18~[0:60]		&	$-$5~[$-$6.5:$-$4.5]	&	0.005~[0.0096:0.0679]	&	20~[10:20]		&	1.0~[1.05:1.15]		&	$-$0.5~[$-$1.5:$-$0.5]	&	1~[1:1000] \\
ISO-Oph~193		&	75$...$121	&	18~[0:53]		&	$-$4.5~[$-$7:$-$5]		&	0.005~[0.005:0.0679]	&	15~[10:20]		&	1.0~[1.05:1.15]		&	$-$0.5~[$-$1.5:$-$0.5]	&	1~[1:1000] \\
CFHTWIR-Oph~66	&	25$...$48	&	87~[84:90]		&	$-$6.5~[$-$7:$-$6]		&	0.0679~[0.0096:0.2500]	&	15~[15:20]		&	1.2~[1.1:1.2]		&	$-$1~[$-$1.5:$-$0.5]	&	1000~[1000:1000] \\
GY92~3			&	20$...$40	&	41~[0.0:53]		&	$-$5.5~[$-$6:$-$4]		&	0.0184~[0.0096:0.0354]	&	20~[15:20]		&	1.05~[1.0:1.1]		&	$-$1~[$-$1.5:$-$0.5]	&	1~[1:1000] \\
GY92~264		&	20$...$40	&	32~[0.0:53]		&	$-$6~[$-$6.5:$-$4]		&	0.0184~[0.0184:0.0354]	&	20~[15:20]		&	1.1~[1.0:1.15]		&	$-$1.5~[$-$1.5:$-$0.5]	&	1~[1:1000] \\
GY92~310		&	15$...$30	&	32~[0.0:46]		&	$-$5~[$-$5:$-$4]		&	0.0184~[0.005:0.1303]	&	20~[15:20]		&	1.1~[1.1:1.1]		&	$-$1.5~[$-$1.5:$-$1.5]	&	1~[1:1000] \\
\hline
\end{tabular}
\tablefoot{
\tablefoottext{a}{Masses were derived by converting the spectral types from Table~\ref{table1} to temperature using the empirical relation from \citet{Luhman2003}, and the Dusty evolutionary models \citep{Chabrier2000}. The range in masses denotes a typical error in spectral classification of half a sub-spectral type.}
\tablefoottext{b}{In MCFOST, the scale height and the dust surface density are treated as power-law distributions of the form $h(r)=h_{0}(r/r_{0})^{\beta}$ and $\Sigma(r)=\Sigma_{0}(r/r_{0})^{-p}$, respectively, where $r$ is the radial coordinate in the disk plane, and $h_{0}$ is the scale height at a fiducial radius at 100 AU.}
}
\end{table*}

\subsection{Disk modelling}
Disks around brown dwarfs radiate from the infrared to the millimeter range. Their geometry and dust properties affect the amount of radiation that is absorbed or scattered, setting the temperature of the dust and the amount of re-emitted thermal radiation and shaping their SEDs. We compared the observed brown dwarf SEDs to the same grid of disk models as presented by \citet{Harvey2012b}, built with the radiative transfer code using the Monte-Carlo method MCFOST \citep{Pinte2006,Pinte2009}. Our goal is to study the scale height and flaring angle of the disk. We summarise here the modelling setup in building the grid of disk models, a more detailed description is given in \citet{Harvey2012b}. 

In MCFOST, the disk extends from an inner radius ($r_{in}$) to an outer limit radius ($r_{out}$) and is parametrized as follows: the dust density distribution has a Gaussian vertical profile, while the dust surface density and the scale height are power-law distributions ($\rho(r,z)=\rho_0(r)\,\exp(-z^2/2\,h^2(r))$, $\Sigma(r) = \Sigma_0\,(r/r_0)^{-p}$ and $ h(r) = h_0\, (r/r_0)^{\gamma}$, respectively, where $r$ is the radial coordinate in the equatorial plane and $h_0$ is the scale height at a fiducial radius $r_0 = $ 100 AU). The dust grains are defined as homogeneous and spherical  with a differential grain size distribution of the form $dn(a) \propto a^{-3.5} da$, between a minimum grain size of 0.03~$\mu$m and a maximum grain size between  1~$\mu$m and 1~mm. Dust extinction and scattering opacities, scattering phase functions, and Mueller matrices are calculated using Mie theory. Furthermore, the following assumptions are taken: the grain properties are assumed to be independent of position within the disk, the grain size distribution and dust composition are fixed parameters, and the outer radius is fixed at 100~AU. The parameters left free to vary in the grid are the inner disk radius and disk inclination, the scale-height parameters, the surface density exponent, and the maximum grain size. 

In the comparison of the observed SEDs with the grid of disk models, the known stellar parameters are used as input (effective temperature, visual extinction), and represented by the respective NextGen atmospheric model \citep{Hauschildt1999}. A Bayesian analysis is used to determine the most likely range of parameters that reproduce the observations. Figure~\ref{sed} shows one of the most likely fitting models overlaid on the brown dwarfs' observed SEDs (blue solid line). Table~\ref{table3} presents the resulting disk parameters of the synthetic SED that match the observations, found using a $\chi$$^{2}$ minimisation. We also give the ranges of validity for each parameter quoted in brackets derived from the Bayesian inference, corresponding to regions where P~$>$~0.5~$\times$~P$_{Max}$. We have not modelled the disk of GY92~344 since its particular geometry very likely introduces in the photometric measurements strong contamination from scattered light, and also makes the measurement of extinction towards the object very uncertain. In this disk orientation, the SED is very sensitive to the inclination angle, and the coarse sampling in our grid of models would be insufficient to derive a reliable model.

\begin{figure*}[ht]
   \centering
 \includegraphics[width=\linewidth]{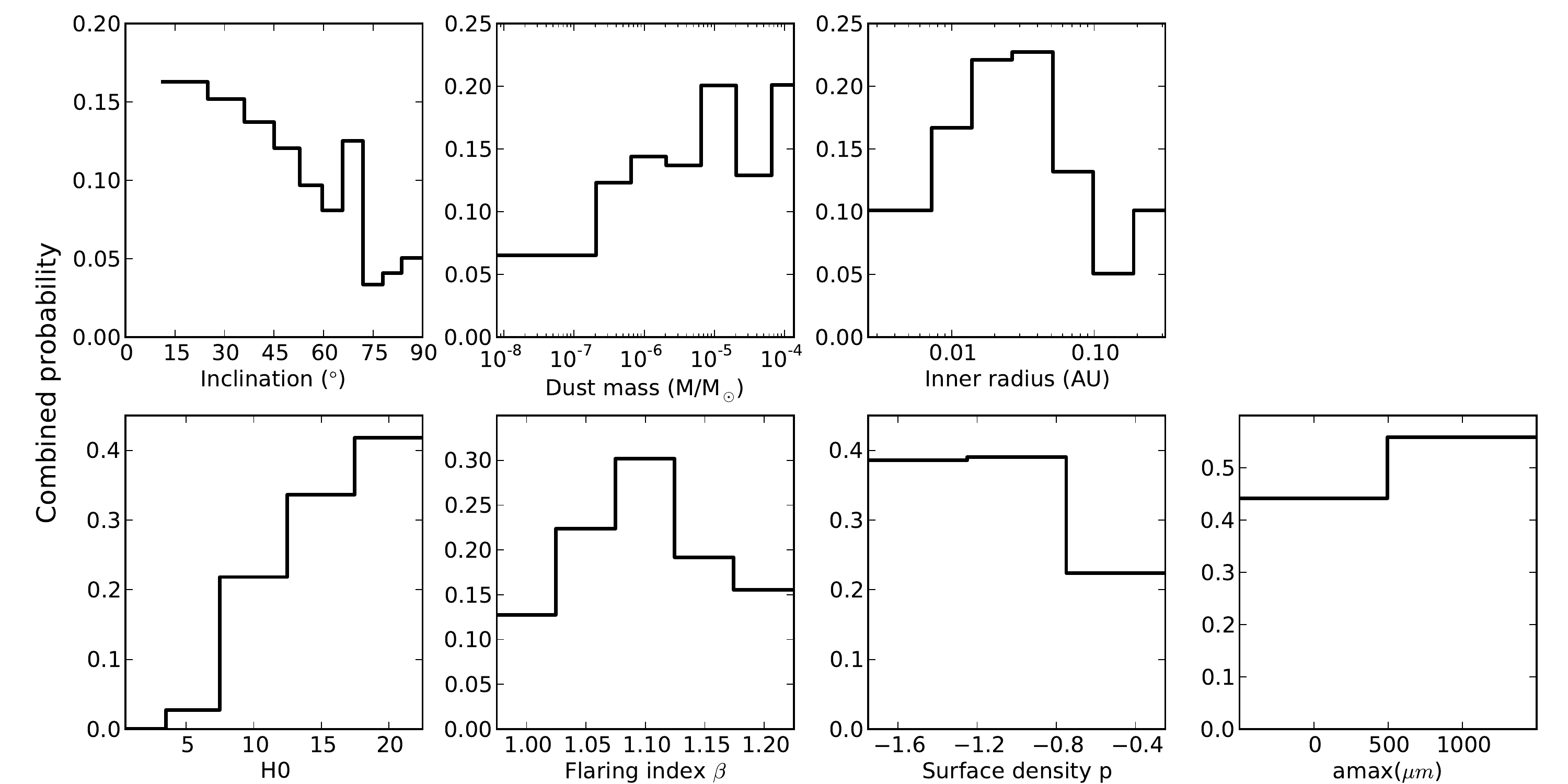}
\caption{Combined probability distributions for the disk parameters of the 11 brown dwarf disks modelled.}
\label{prob}
\end{figure*}

%______________________________________________________________
\section{Discussion} \label{discussion}
With the new \emph{Herschel} observations, the combination of the far-IR measurements with previous data at shorter wavelengths, and for one object ALMA millimetre observations, results in an unprecedented dataset for the $\rho$~Ophiuchi brown dwarfs. Taking advantage of the extended SED coverage, we attempted for the first time to characterise their disk structural properties by using the radiative transfer code MCFOST. The histograms of the combined probability distributions for the 11 brown dwarfs modelled are shown in Fig.~\ref{prob} for each free parameter explored in the MCFOST grid.

\subsection{Structural properties of the brown dwarf disks}\label{disc1}
The disks are found at a variety of inclinations and do not show a particular alignment on sky, though for most objects this parameter is only loosely constrained (see the ranges of validity in Table~\ref{table3}). The SED of CFHTWIR-Oph~66 is consistent with a nearly edge-on disk geometry (87${\degr}$, range of validity between 84${\degr}$ and 90${\degr}$). This object has previously been characterised by \citet{AlvesdeOliveira2008} as being variable in the near-IR, with variations consistent with a change in extinction likely originating from orbiting material in the circumstellar environment. This scenario is further supported by the disk orientation derived here.

The histogram of the combined probability distribution of the inner disk radius for all brown dwarfs shows a narrow range from where we calculate that they can be described by an inner disk radius between 0.01 and 0.07~AU. This parameter is also well constrained for all objects individually. The near and mid-IR data are sensitive to the inner disk, but the extension of the SED characterisation into the far-IR with \emph{Herschel} is crucial in constraining other disk parameters and therefore improving the model solution. Within the sample of 11 brown dwarfs, we searched for a possible correlation between the inner radius and the mass of the central object, similar to what was obtained for young Herbig Ae and late Be stars \citep{Millan-Gabet2007}, but we did not find any clear trend.

In the wavelength range of our observations, typically $\le$70~$\mu$m, the disk emission is still optically thick, and therefore any attempt to derive dust masses from the disk emission at these wavelengths is likely to be askew. Assuming a gas-to-dust ratio of 100, the derived disk masses are between 0.001 and 0.01~M$_{\sun}$ for the brown dwarfs with an M6 spectral type (corresponding to masses between 0.075 and 0.121~M$_{\sun}$ for the central object), and systematically lower, between $\sim$3$\times10^{-5}$ and 0.001~M$_{\sun}$ for the four brown dwarfs with spectral types around M8 (brown dwarf masses between 0.015 and 0.048~M$_{\sun}$). The only exception is GY92~371, that has a spectral type of M6 and a low disk mass of $\sim$3$\times10^{-5}$ (but with a large validity range). This object corresponds to the only binary known in our sample \citep{Ratzka2005} with a companion at a distance of 0.347$\pm$0.001$\arcsec$. At this separation, the binary cannot be resolved in the mid or far-IR observations. Since our grid of models assumes all central sources to be single, the derived disk parameters could be affected. For the remaining sources, the values are consistent with a disk mass-to-central object ratio of less than a few percent. However, the exact result can only be derived by observations at longer wavelengths, since our data are not sensitive to the outer part of the disk. The only source for which we can accurately study the disk mass using the grid of models is GY92~204, where the SED is complete up until the millimetre regime (Sect.~\ref{disc3}).

The \emph{Herschel} measurements at 70~$\mu$m allow us to constrain the flaring index ($\beta$) of the brown dwarf disks that shows a narrow peaked probability distribution for each individual object, and for the combined sample as well. We find a range of validity for the histogram of the combined probability between 1.05 and 1.2, consistent with a flared disk geometry. Furthermore, we can exclude values of the scale height smaller than 10~AU (measured at a fiducial radius of 100~AU), though we cannot exclude values larger than $\sim$20~AU, since our grid does not probe this parameter beyond this range. 

Finally, neither the surface density index (p) nor the maximum grain size (amax) are constrained by our observations.

\subsection{Comparison to other disk studies}

Studies of brown dwarf disks in the far-IR regime are still relatively scarce. Our results can be directly compared to those presented in \citet{Harvey2012b,Harvey2012a}, because we used the same instrument and, more importantly, the same grid of models. The first immediate conclusion is that the analysis of the 11 brown dwarf disks in $\rho$~Ophiuchi reveals a much narrower combined probability distribution of the inner disk radius and the flaring index than that in \citet{Harvey2012b}. These parameters are well constrained for each individual brown dwarf, and the result is consistent with the fact that our sample is restricted to coeval objects that span a limited range in mass. For the same parameters, Fig.~10 of \citet{Harvey2012b} shows a large spread of values. This is likely to be a reflection of the range in ages (from $\sim$1 to 10~Myr) and larger differences in central mass of their entire sample taken over several clusters. Unfortunately, the authors have not published the results of the modelling for particular subsets of objects with similar ages or masses, thus we cannot extend our comparative analysis. We can however emphasise the point that based on those two disk structure parameters (inner radii and flaring index), which are well constrained by our observations, our coeval sample of brown dwarfs in $\rho$~Ophiuchi presents very similar properties. This implies that all these brown dwarf disks have evolved, and probably formed, in similar circumstances. 

In another recent study using \emph{Herschel} observations, \citet{Spezzi2013} have used a different radiative transfer code to study the disk properties of low-mass stars in the $\sim$2Myr old Chamaeleon~II cluster. Despite the fact that their sample does not reach into the substellar regime, a few of their targets have spectral types corresponding to low masses of $\sim$0.3~M$_{\sun}$. For these objects, they find a typical inner radius between 0.02 and 0.1~AU and a flaring index between 1.1 and 1.2. These results are consistent with our findings, and suggest that disk properties change smoothly from brown dwarfs to low mass stars.  

An attempt by \citet{Riaz2012} to detect brown dwarf disks with \emph{Herschel} in the TW~Hydrae association ($\sim$10~Myr) has not been successful. Based on the derived upper limits at far-IR wavelengths and radiative transfer modelling, they estimate that the TW~Hydrae brown dwarf disks have a lower mass than younger counterparts. Given that our longest wavelength observations are likely not sensitive to the outer part of the disk, we cannot test this result. 

Finally, it is relevant to compare our results to the brown dwarf 2MASS~J04442713+2512164, a member of Taurus (M7.25, 1$-$3~Myr) with an extensive SED coverage that was studied thoroughly by \citet{Bouy2008} using MCFOST. They find that the well constrained disk structural properties are the inner disk radius with values between 0.02 and 0.1~AU and a peak probability at approximately 0.04~AU, the flaring index with a value between 1.1 and 1.2, and the scale height which is found to be in the range of 30 to 60~AU, with a most probable value around 45~AU. The high value derived for the scale height is interpreted as being consistent with the hydrostatic scale height predicted for brown dwarfs by \citet{Walker2004}. The result for this young brown dwarf is in good agreement with results in $\rho$~Ophiuchi. Although we cannot constrain the scale height of our brown dwarf disks, we have excluded values below 10~AU. Together, these findings seem to contradict the indications that disks around brown dwarfs are significantly flatter than for disks surrounding T~Tauri stars \citep[e.g.,][]{Szucs2010}. In fact, this interpretation has recently been revisited by \citet{Mulders2012}. Using radiative transfer modelling of median SEDs of T~Tauri stars and brown dwarfs separately, they conclude that the reduction in the degree of flaring observed in brown dwarfs disks does not imply a structural difference in the disks but instead can be explained by the fact that for lower mass objects, the disk emission at a particular wavelength corresponds to different radii, giving rise to the differences in the measured scale heights \citep[see also][]{Apai2013}. Our observations seem to corroborate these findings.  

\subsection{Combining Herschel and ALMA observations: GY92~204}
\label{disc3}

We compared the results of our SED modelling of \object{GY92~204} (\object{ISO$-$Oph~102}) combining the \emph{Herschel} and ALMA observations, with those presented by \citet{Ricci2012}. We adopt for this object a spectral type of M6 \citep{Natta2002}, although this classification has been contested by \citet{Wilking2005} who classify it as an M5.5. We converted the spectral type to temperature using the empirical relation from \citet{Luhman2003}, and used the Dusty evolutionary models \citep{Chabrier2000} to derive the expected mass from its effective temperature. Taking into account an error in spectral classification of half a sub-spectral type, the mass range predicted by the evolutionary models is between 75 and 121~M$_{Jup}$, placing this object at the substellar limit. 

The most probable fit from our model grid gives an inclination around 63${\degr}$, an inner radius of 0.035~AU, a scale height of 15~AU with a flaring index $\beta\sim$1.15, an exponent for dust settling of $-$1.5, and a dust mass of 1$\times10^{-5}$~$M_{\sun}$. If we assume a gas-to-dust mass ratio of 100, this corresponds to a disk mass of 0.001~$M_{\sun}$ ($\sim$1~M$_{Jup}$), and a fraction of 0.8 to 1.3$\%$ of the mass of the central object (75 to 121~M$_{Jup}$). \citet{Ricci2012} find a disk mass that varies between $\sim$2 to 6$\times10^{-6}$~$M_{\sun}$ or $\sim$2 to 6$\times10^{-4}$~$M_{\sun}$ depending on the assumed dust opacity, which is their most uncertain parameter.  In MCFOST, the opacity law is not assumed directly, and opacities are calculated using Mie theory from the grain size distribution and dust composition. For reference, for the modelling of GY92~204, we calculate $\kappa$$^{dust}_{0.89mm}$~=~0.37~cm$^{2}$/g for amax = 1~$\mu$m and $\kappa$$^{dust}_{0.89mm}$~=~3.6~cm$^{2}$/g for amax = 1~mm. 

The comparison between results achieved with different modelling approaches is not straightforward. We conclude that our disk mass estimate is approximately in agreement with that derived by \citet{Ricci2012}, assuming higher dust opacity values. Furthermore, the disk-to-central object mass ratio is consistent with what is found for other T~Tauri stars in $\rho$~Oph \citep{Andrews2009,Andrews2010} and other regions \citep[e.g.][]{Williams2011,Andrews2013}.

%______________________________________________________________

\section{Conclusions} \label{conclusions}
We present the first \emph{Herschel} survey of disks around brown dwarfs of an entire star forming region, $\rho$~Ophiuchi. We detected 12 brown dwarf disks at 70~$\mu$m, out of which 11/3 are also detected at 100/160~$\mu$m. Combining the new far-IR data with information from the literature, we constructed SEDs and used the MCFOST radiative transfer code to investigate the likely range of parameters that reproduce the structural properties of the disks of 11 sources.

We find that the brown dwarfs in the $\sim$1~Myr old $\rho$~Ophiuchi cluster are surrounded by disks well described by an inner radius between 0.01 and 0.07~AU. Furthermore, the \emph{Herschel} measurements at 70~$\mu$m can constrain the degree of flaring of the disks. We find that they are well described by a flaring index between 1.05 and 1.2, and we can exclude values of the scale height lower than 10~AU (measured at a fiducial radius of 100~AU). This suggests that brown dwarf disks share a similar degree of flaring as T~Tauri stars. Furthermore, the disk parameters constrained by our SED coverage (inner radius and flaring index), describe well all the 11 brown dwarf disks, indicating that for these coeval sample of a small range in masses, objects have a similar disk evolution and, perhaps, formation. We note, however, that due to the \emph{Herschel} sensitivity limit, our observations are incomplete, being biased towards detecting the brighter disks. Thus, our findings on the disk parameters cannot be extrapolated to the entire substellar population of this cluster.

We combined the new \emph{Herschel} photometry with recently published ALMA submillimetre observations for one of the brown dwarfs with an estimated mass at the substellar limit (\object{GY92-204}), and derived the most probable parameters by comparing it to the same grid of models produced with MCFOST. We estimate a disk mass of 0.001~$M_{\sun}$ ($\sim$1~M$_{Jup}$). According to the evolutionary models, this brown dwarf has a mass between 75 and 121~M$_{Jup}$, meaning that the disk mass is 0.8 to 1.3$\%$  of the mass of the central object. This result is consistent with the findings from ALMA observations, and also with the disk-to-star mass ratio found for T~Tauri stars. 

Future ALMA measurements of more brown dwarfs disks, combined with \emph{Herschel} data which we have shown can constrain parameters of the structure geometry, will certainly provide a better understanding of disks around brown dwarfs. 

%______________________________________________________________

\begin{acknowledgements}
We thank the referee for a helpful review. This work was partly supported by the Hungarian national Research Fund OTKA K101393. C.K.'s work has been supported by the PECS-98073 grant of the European Space Agency and the Hungarian Space Office, the K-104607 grant of the Hungarian Research Fund and the Bolyai Fellowship of the Hungarian Academy of Sciences. C. Pinte acknowledges funding from the European Commission's 7$^\mathrm{th}$ Framework Program (contract PERG06-GA-2009-256513) and from Agence Nationale pour la Recherche (ANR) of France under contract ANR-2010-JCJC-0504-01. Calculations were performed at Service Commun de Calcul Intensif de l'Observatoire de Grenoble (SCCI) on the fostino super-computer funded by ANR (contracts ANR-07-BLAN-0221, ANR-2010-JCJC-0504-01 and ANR-2010-JCJC-0501-01) and  the European Commission's FP7 (contract PERG06-GA-2009-256513). PACS has been developed by a consortium of institutes led by MPE (Germany) and including UVIE (Austria); KU Leuven, CSL, IMEC (Belgium); CEA, LAM (France); MPIA (Germany); INAF- IFSI/OAA/OAP/OAT, LENS, SISSA (Italy); IAC (Spain). This development has been supported by the funding agencies BMVIT (Austria), ESA-PRODEX (Belgium), CEA/CNES (France), DLR (Germany), ASI/INAF (Italy), and CICYT/MCYT (Spain). 
\end{acknowledgements}

\bibliographystyle{aa} 
\bibliography{ophdiscs}

\begin{thebibliography}{75}
\expandafter\ifx\csname natexlab\endcsname\relax\def\natexlab#1{#1}\fi

\bibitem[{{Alves de Oliveira} \& {Casali}(2008)}]{AlvesdeOliveira2008}
{Alves de Oliveira}, C. \& {Casali}, M. 2008, \aap, 485, 155

\bibitem[{{Alves de Oliveira} {et~al.}(2012){Alves de Oliveira}, {Moraux},
  {Bouvier}, \& {Bouy}}]{AlvesdeOliveira2012}
{Alves de Oliveira}, C., {Moraux}, E., {Bouvier}, J., \& {Bouy}, H. 2012, \aap,
  539, A151

\bibitem[{{Alves de Oliveira} {et~al.}(2010){Alves de Oliveira}, {Moraux},
  {Bouvier}, {Bouy}, {Marmo}, \& {Albert}}]{AlvesdeOliveira2010}
{Alves de Oliveira}, C., {Moraux}, E., {Bouvier}, J., {et~al.} 2010, \aap, 515,
  A75

\bibitem[{{Andr{\'e}} {et~al.}(2010){Andr{\'e}}, {Men'shchikov}, {Bontemps},
  {K{\"o}nyves}, {Motte}, {Schneider}, {Didelon}, {Minier}, {Saraceno},
  {Ward-Thompson}, {di Francesco}, {White}, {Molinari}, {Testi}, {Abergel},
  {Griffin}, {Henning}, {Royer}, {Mer{\'{\i}}n}, {Vavrek}, {Attard},
  {Arzoumanian}, {Wilson}, {Ade}, {Aussel}, {Baluteau}, {Benedettini},
  {Bernard}, {Blommaert}, {Cambr{\'e}sy}, {Cox}, {di Giorgio}, {Hargrave},
  {Hennemann}, {Huang}, {Kirk}, {Krause}, {Launhardt}, {Leeks}, {Le Pennec},
  {Li}, {Martin}, {Maury}, {Olofsson}, {Omont}, {Peretto}, {Pezzuto}, {Prusti},
  {Roussel}, {Russeil}, {Sauvage}, {Sibthorpe}, {Sicilia-Aguilar}, {Spinoglio},
  {Waelkens}, {Woodcraft}, \& {Zavagno}}]{Andre2010}
{Andr{\'e}}, P., {Men'shchikov}, A., {Bontemps}, S., {et~al.} 2010, \aap, 518,
  L102

\bibitem[{{Andr{\'e}} {et~al.}(2012){Andr{\'e}}, {Ward-Thompson}, \&
  {Greaves}}]{Andre2012}
{Andr{\'e}}, P., {Ward-Thompson}, D., \& {Greaves}, J. 2012, Science, 337, 69

\bibitem[{{Andrews} {et~al.}(2013){Andrews}, {Rosenfeld}, {Kraus}, \&
  {Wilner}}]{Andrews2013}
{Andrews}, S.~M., {Rosenfeld}, K.~A., {Kraus}, A.~L., \& {Wilner}, D.~J. 2013,
  \apj, 771, 129

\bibitem[{{Andrews} {et~al.}(2009){Andrews}, {Wilner}, {Hughes}, {Qi}, \&
  {Dullemond}}]{Andrews2009}
{Andrews}, S.~M., {Wilner}, D.~J., {Hughes}, A.~M., {Qi}, C., \& {Dullemond},
  C.~P. 2009, \apj, 700, 1502

\bibitem[{{Andrews} {et~al.}(2010){Andrews}, {Wilner}, {Hughes}, {Qi}, \&
  {Dullemond}}]{Andrews2010}
{Andrews}, S.~M., {Wilner}, D.~J., {Hughes}, A.~M., {Qi}, C., \& {Dullemond},
  C.~P. 2010, \apj, 723, 1241

\bibitem[{{Apai}(2013)}]{Apai2013}
{Apai}, D. 2013, Astronomische Nachrichten, 334, 57

\bibitem[{{Baraffe} {et~al.}(1998){Baraffe}, {Chabrier}, {Allard}, \&
  {Hauschildt}}]{Baraffe1998}
{Baraffe}, I., {Chabrier}, G., {Allard}, F., \& {Hauschildt}, P.~H. 1998, \aap,
  337, 403

\bibitem[{{Barsony} {et~al.}(2005){Barsony}, {Ressler}, \&
  {Marsh}}]{Barsony2005}
{Barsony}, M., {Ressler}, M.~E., \& {Marsh}, K.~A. 2005, \apj, 630, 381

\bibitem[{{Basu} \& {Vorobyov}(2012)}]{Basu2012}
{Basu}, S. \& {Vorobyov}, E.~I. 2012, \apj, 750, 30

\bibitem[{{Bontemps} {et~al.}(2001){Bontemps}, {Andr{\'e}}, {Kaas}, {Nordh},
  {Olofsson}, {Huldtgren}, {Abergel}, {Blommaert}, {Boulanger}, {Burgdorf},
  {Cesarsky}, {Cesarsky}, {Copet}, {Davies}, {Falgarone}, {Lagache},
  {Montmerle}, {P{\'e}rault}, {Persi}, {Prusti}, {Puget}, \&
  {Sibille}}]{Bontemps2001}
{Bontemps}, S., {Andr{\'e}}, P., {Kaas}, A.~A., {et~al.} 2001, \aap, 372, 173

\bibitem[{{Bouy} {et~al.}(2008){Bouy}, {Hu{\'e}lamo}, {Pinte}, {Olofsson},
  {Barrado Y Navascu{\'e}s}, {Mart{\'{\i}}n}, {Pantin}, {Monin}, {Basri},
  {Augereau}, {M{\'e}nard}, {Duvert}, {Duch{\^e}ne}, {Marchis}, {Bayo},
  {Bottinelli}, {Lefort}, \& {Guieu}}]{Bouy2008}
{Bouy}, H., {Hu{\'e}lamo}, N., {Pinte}, C., {et~al.} 2008, \aap, 486, 877

\bibitem[{{Chabrier} {et~al.}(2000){Chabrier}, {Baraffe}, {Allard}, \&
  {Hauschildt}}]{Chabrier2000}
{Chabrier}, G., {Baraffe}, I., {Allard}, F., \& {Hauschildt}, P. 2000, \apj,
  542, 464

\bibitem[{{Comer{\'o}n} {et~al.}(2010){Comer{\'o}n}, {Testi}, \&
  {Natta}}]{Comeron2010}
{Comer{\'o}n}, F., {Testi}, L., \& {Natta}, A. 2010, \aap, 522, A47+

\bibitem[{{Cushing} {et~al.}(2000){Cushing}, {Tokunaga}, \&
  {Kobayashi}}]{Cushing2000}
{Cushing}, M.~C., {Tokunaga}, A.~T., \& {Kobayashi}, N. 2000, \aj, 119, 3019

\bibitem[{{Cutri} {et~al.}(2003){Cutri}, {Skrutskie}, {van Dyk}, {Beichman},
  {Carpenter}, {Chester}, {Cambresy}, {Evans}, {Fowler}, {Gizis}, {Howard},
  {Huchra}, {Jarrett}, {Kopan}, {Kirkpatrick}, {Light}, {Marsh}, {McCallon},
  {Schneider}, {Stiening}, {Sykes}, {Weinberg}, {Wheaton}, {Wheelock}, \&
  {Zacarias}}]{Cutri2003}
{Cutri}, R.~M., {Skrutskie}, M.~F., {van Dyk}, S., {et~al.} 2003, VizieR Online
  Data Catalog, 2246, 0

\bibitem[{{Evans} {et~al.}(2009){Evans}, {Dunham}, {J{\o}rgensen}, {Enoch},
  {Mer{\'{\i}}n}, {van Dishoeck}, {Alcal{\'a}}, {Myers}, {Stapelfeldt},
  {Huard}, {Allen}, {Harvey}, {van Kempen}, {Blake}, {Koerner}, {Mundy},
  {Padgett}, \& {Sargent}}]{Evans2009}
{Evans}, II, N.~J., {Dunham}, M.~M., {J{\o}rgensen}, J.~K., {et~al.} 2009,
  \apjs, 181, 321

\bibitem[{{Geers} {et~al.}(2011){Geers}, {Scholz}, {Jayawardhana}, {Lee},
  {Lafreni{\`e}re}, \& {Tamura}}]{Geers2011}
{Geers}, V., {Scholz}, A., {Jayawardhana}, R., {et~al.} 2011, \apj, 726, 23

\bibitem[{{Greene} \& {Young}(1992)}]{Greene1992}
{Greene}, T.~P. \& {Young}, E.~T. 1992, \apj, 395, 516

\bibitem[{{Griffin} {et~al.}(2010){Griffin}, {Abergel}, {Abreu}, {Ade},
  {Andr{\'e}}, {Augueres}, {Babbedge}, {Bae}, {Baillie}, {Baluteau}, {Barlow},
  {Bendo}, {Benielli}, {Bock}, {Bonhomme}, {Brisbin}, {Brockley-Blatt},
  {Caldwell}, {Cara}, {Castro-Rodriguez}, {Cerulli}, {Chanial}, {Chen},
  {Clark}, {Clements}, {Clerc}, {Coker}, {Communal}, {Conversi}, {Cox},
  {Crumb}, {Cunningham}, {Daly}, {Davis}, {de Antoni}, {Delderfield}, {Devin},
  {di Giorgio}, {Didschuns}, {Dohlen}, {Donati}, {Dowell}, {Dowell}, {Duband},
  {Dumaye}, {Emery}, {Ferlet}, {Ferrand}, {Fontignie}, {Fox}, {Franceschini},
  {Frerking}, {Fulton}, {Garcia}, {Gastaud}, {Gear}, {Glenn}, {Goizel},
  {Griffin}, {Grundy}, {Guest}, {Guillemet}, {Hargrave}, {Harwit}, {Hastings},
  {Hatziminaoglou}, {Herman}, {Hinde}, {Hristov}, {Huang}, {Imhof}, {Isaak},
  {Israelsson}, {Ivison}, {Jennings}, {Kiernan}, {King}, {Lange}, {Latter},
  {Laurent}, {Laurent}, {Leeks}, {Lellouch}, {Levenson}, {Li}, {Li},
  {Lilienthal}, {Lim}, {Liu}, {Lu}, {Madden}, {Mainetti}, {Marliani}, {McKay},
  {Mercier}, {Molinari}, {Morris}, {Moseley}, {Mulder}, {Mur}, {Naylor},
  {Nguyen}, {O'Halloran}, {Oliver}, {Olofsson}, {Olofsson}, {Orfei}, {Page},
  {Pain}, {Panuzzo}, {Papageorgiou}, {Parks}, {Parr-Burman}, {Pearce},
  {Pearson}, {P{\'e}rez-Fournon}, {Pinsard}, {Pisano}, {Podosek}, {Pohlen},
  {Polehampton}, {Pouliquen}, {Rigopoulou}, {Rizzo}, {Roseboom}, {Roussel},
  {Rowan-Robinson}, {Rownd}, {Saraceno}, {Sauvage}, {Savage}, {Savini},
  {Sawyer}, {Scharmberg}, {Schmitt}, {Schneider}, {Schulz}, {Schwartz},
  {Shafer}, {Shupe}, {Sibthorpe}, {Sidher}, {Smith}, {Smith}, {Smith},
  {Spencer}, {Stobie}, {Sudiwala}, {Sukhatme}, {Surace}, {Stevens}, {Swinyard},
  {Trichas}, {Tourette}, {Triou}, {Tseng}, {Tucker}, {Turner}, {Vaccari},
  {Valtchanov}, {Vigroux}, {Virique}, {Voellmer}, {Walker}, {Ward}, {Waskett},
  {Weilert}, {Wesson}, {White}, {Whitehouse}, {Wilson}, {Winter}, {Woodcraft},
  {Wright}, {Xu}, {Zavagno}, {Zemcov}, {Zhang}, \& {Zonca}}]{Griffin2010}
{Griffin}, M.~J., {Abergel}, A., {Abreu}, A., {et~al.} 2010, \aap, 518, L3

\bibitem[{{Han} {et~al.}(2013){Han}, {Jung}, {Udalski}, {Sumi}, {Gaudi},
  {Gould}, {Bennett}, {Tsapras}, {Szyma{\'n}ski}, {Kubiak}, {Pietrzy{\'n}ski},
  {Soszy{\'n}ski}, {Skowron}, {Koz{\l}owski}, {Poleski}, {Ulaczyk},
  {Wyrzykowski}, {Pietrukowicz}, {Abe}, {Bond}, {Botzler}, {Chote}, {Freeman},
  {Fukui}, {Furusawa}, {Harris}, {Itow}, {Ling}, {Masuda}, {Matsubara},
  {Muraki}, {Ohnishi}, {Rattenbury}, {Saito}, {Sullivan}, {Sweatman}, {Suzuki},
  {Tristram}, {Wada}, {Yock}, {Batista}, {Christie}, {Choi}, {DePoy}, {Dong},
  {Hwang}, {Kavka}, {Lee}, {Monard}, {Natusch}, {Ngan}, {Park}, {Pogge},
  {Porritt}, {Shin}, {Tan}, {Yee}, {Alsubai}, {Bozza}, {Bramich}, {Browne},
  {Dominik}, {Horne}, {Hundertmark}, {Ipatov}, {Kains}, {Liebig}, {Snodgrass},
  {Steele}, \& {Street}}]{Han2013}
{Han}, C., {Jung}, Y.~K., {Udalski}, A., {et~al.} 2013, ArXiv e-prints

\bibitem[{{Harvey} {et~al.}(2012{\natexlab{a}}){Harvey}, {Henning}, {Liu},
  {M{\'e}nard}, {Pinte}, {Wolf}, {Cieza}, {Evans}, \& {Pascucci}}]{Harvey2012b}
{Harvey}, P.~M., {Henning}, T., {Liu}, Y., {et~al.} 2012{\natexlab{a}}, \apj,
  755, 67

\bibitem[{{Harvey} {et~al.}(2012{\natexlab{b}}){Harvey}, {Henning},
  {M{\'e}nard}, {Wolf}, {Liu}, {Cieza}, {Evans}, {Pascucci}, {Mer{\'{\i}}n}, \&
  {Pinte}}]{Harvey2012a}
{Harvey}, P.~M., {Henning}, T., {M{\'e}nard}, F., {et~al.} 2012{\natexlab{b}},
  \apjl, 744, L1

\bibitem[{{Hauschildt} {et~al.}(1999){Hauschildt}, {Allard}, \&
  {Baron}}]{Hauschildt1999}
{Hauschildt}, P.~H., {Allard}, F., \& {Baron}, E. 1999, \apj, 512, 377

\bibitem[{{Hayashi} \& {Nakano}(1963)}]{Hayashi1963}
{Hayashi}, C. \& {Nakano}, T. 1963, Progress of Theoretical Physics, 30, 460

\bibitem[{{Hennebelle} \& {Chabrier}(2008)}]{Hennebelle2008}
{Hennebelle}, P. \& {Chabrier}, G. 2008, \apj, 684, 395

\bibitem[{{Hodapp} {et~al.}(2004){Hodapp}, {Walker}, {Reipurth}, {Wood},
  {Bally}, {Whitney}, \& {Connelley}}]{Hodapp2004}
{Hodapp}, K.~W., {Walker}, C.~H., {Reipurth}, B., {et~al.} 2004, \apjl, 601,
  L79

\bibitem[{{Joergens} {et~al.}(2012){Joergens}, {Pohl}, {Sicilia-Aguilar}, \&
  {Henning}}]{Joergens2012}
{Joergens}, V., {Pohl}, A., {Sicilia-Aguilar}, A., \& {Henning}, T. 2012, \aap,
  543, A151

\bibitem[{{Kroupa} \& {Bouvier}(2003)}]{Kroupa2003}
{Kroupa}, P. \& {Bouvier}, J. 2003, \mnras, 346, 369

\bibitem[{{Kubas} {et~al.}(2012){Kubas}, {Beaulieu}, {Bennett}, {Cassan},
  {Cole}, {Lunine}, {Marquette}, {Dong}, {Gould}, {Sumi}, {Batista},
  {Fouqu{\'e}}, {Brillant}, {Dieters}, {Coutures}, {Greenhill}, {Bond},
  {Nagayama}, {Udalski}, {Pompei}, {N{\"u}rnberger}, \& {Le
  Bouquin}}]{Kubas2012}
{Kubas}, D., {Beaulieu}, J.~P., {Bennett}, D.~P., {et~al.} 2012, \aap, 540, A78

\bibitem[{{Kumar}(1963)}]{Kumar1963}
{Kumar}, S.~S. 1963, \apj, 137, 1121

\bibitem[{{Luhman} {et~al.}(2007){Luhman}, {Joergens}, {Lada}, {Muzerolle},
  {Pascucci}, \& {White}}]{Luhman2007}
{Luhman}, K.~L., {Joergens}, V., {Lada}, C., {et~al.} 2007, Protostars and
  Planets V, 443

\bibitem[{{Luhman} {et~al.}(1997){Luhman}, {Liebert}, \& {Rieke}}]{Luhman1997}
{Luhman}, K.~L., {Liebert}, J., \& {Rieke}, G.~H. 1997, \apjl, 489, L165+

\bibitem[{{Luhman} \& {Rieke}(1999)}]{Luhman1999}
{Luhman}, K.~L. \& {Rieke}, G.~H. 1999, \apj, 525, 440

\bibitem[{{Luhman} {et~al.}(2003){Luhman}, {Stauffer}, {Muench}, {Rieke},
  {Lada}, {Bouvier}, \& {Lada}}]{Luhman2003}
{Luhman}, K.~L., {Stauffer}, J.~R., {Muench}, A.~A., {et~al.} 2003, \apj, 593,
  1093

\bibitem[{{McClure} {et~al.}(2010){McClure}, {Furlan}, {Manoj}, {Luhman},
  {Watson}, {Forrest}, {Espaillat}, {Calvet}, {D'Alessio}, {Sargent}, {Tobin},
  \& {Chiang}}]{McClure2010}
{McClure}, M.~K., {Furlan}, E., {Manoj}, P., {et~al.} 2010, \apjs, 188, 75

\bibitem[{{Men'shchikov} {et~al.}(2012){Men'shchikov}, {Andr{\'e}}, {Didelon},
  {Motte}, {Hennemann}, \& {Schneider}}]{Menshchikov2012}
{Men'shchikov}, A., {Andr{\'e}}, P., {Didelon}, P., {et~al.} 2012, \aap, 542,
  A81

\bibitem[{{Millan-Gabet} {et~al.}(2007){Millan-Gabet}, {Malbet}, {Akeson},
  {Leinert}, {Monnier}, \& {Waters}}]{Millan-Gabet2007}
{Millan-Gabet}, R., {Malbet}, F., {Akeson}, R., {et~al.} 2007, Protostars and
  Planets V, 539

\bibitem[{{Mohanty} {et~al.}(2005){Mohanty}, {Jayawardhana}, \&
  {Basri}}]{Mohanty2005}
{Mohanty}, S., {Jayawardhana}, R., \& {Basri}, G. 2005, \apj, 626, 498

\bibitem[{{Muench} {et~al.}(2001){Muench}, {Alves}, {Lada}, \&
  {Lada}}]{Muench2001}
{Muench}, A.~A., {Alves}, J., {Lada}, C.~J., \& {Lada}, E.~A. 2001, \apjl, 558,
  L51

\bibitem[{{Muirhead} {et~al.}(2012){Muirhead}, {Johnson}, {Apps}, {Carter},
  {Morton}, {Fabrycky}, {Pineda}, {Bottom}, {Rojas-Ayala}, {Schlawin},
  {Hamren}, {Covey}, {Crepp}, {Stassun}, {Pepper}, {Hebb}, {Kirby}, {Howard},
  {Isaacson}, {Marcy}, {Levitan}, {Diaz-Santos}, {Armus}, \&
  {Lloyd}}]{Muirhead2012}
{Muirhead}, P.~S., {Johnson}, J.~A., {Apps}, K., {et~al.} 2012, \apj, 747, 144

\bibitem[{{Mulders} \& {Dominik}(2012)}]{Mulders2012}
{Mulders}, G.~D. \& {Dominik}, C. 2012, \aap, 539, A9

\bibitem[{{Mu{\v z}i{\'c}} {et~al.}(2012){Mu{\v z}i{\'c}}, {Scholz}, {Geers},
  {Jayawardhana}, \& {Tamura}}]{Muzic2012}
{Mu{\v z}i{\'c}}, K., {Scholz}, A., {Geers}, V., {Jayawardhana}, R., \&
  {Tamura}, M. 2012, \apj, 744, 134

\bibitem[{{Nakajima} {et~al.}(1995){Nakajima}, {Oppenheimer}, {Kulkarni},
  {Golimowski}, {Matthews}, \& {Durrance}}]{Nakajima1995}
{Nakajima}, T., {Oppenheimer}, B.~R., {Kulkarni}, S.~R., {et~al.} 1995, \nat,
  378, 463

\bibitem[{{Natta} {et~al.}(2002){Natta}, {Testi}, {Comer{\'o}n}, {Oliva},
  {D'Antona}, {Baffa}, {Comoretto}, \& {Gennari}}]{Natta2002}
{Natta}, A., {Testi}, L., {Comer{\'o}n}, F., {et~al.} 2002, \aap, 393, 597

\bibitem[{{Oppenheimer} {et~al.}(1995){Oppenheimer}, {Kulkarni}, {Matthews}, \&
  {Nakajima}}]{Oppenheimer1995}
{Oppenheimer}, B.~R., {Kulkarni}, S.~R., {Matthews}, K., \& {Nakajima}, T.
  1995, Science, 270, 1478

\bibitem[{{Ott}(2010)}]{Ott2010}
{Ott}, S. 2010, in Astronomical Society of the Pacific Conference Series, Vol.
  434, Astronomical Data Analysis Software and Systems XIX, ed. Y.~{Mizumoto},
  K.-I. {Morita}, \& M.~{Ohishi}, 139

\bibitem[{{Padgett} {et~al.}(2008){Padgett}, {Rebull}, {Stapelfeldt},
  {Chapman}, {Lai}, {Mundy}, {Evans}, {Brooke}, {Cieza}, {Spiesman},
  {Noriega-Crespo}, {McCabe}, {Allen}, {Blake}, {Harvey}, {Huard},
  {J{\o}rgensen}, {Koerner}, {Myers}, {Sargent}, {Teuben}, {van Dishoeck},
  {Wahhaj}, \& {Young}}]{Padgett2008}
{Padgett}, D.~L., {Rebull}, L.~M., {Stapelfeldt}, K.~R., {et~al.} 2008, \apj,
  672, 1013

\bibitem[{{Padoan} \& {Nordlund}(2002)}]{Padoan2002}
{Padoan}, P. \& {Nordlund}, {\AA}. 2002, \apj, 576, 870

\bibitem[{{Pilbratt} {et~al.}(2010){Pilbratt}, {Riedinger}, {Passvogel},
  {Crone}, {Doyle}, {Gageur}, {Heras}, {Jewell}, {Metcalfe}, {Ott}, \&
  {Schmidt}}]{Pilbratt2010}
{Pilbratt}, G.~L., {Riedinger}, J.~R., {Passvogel}, T., {et~al.} 2010, \aap,
  518, L1

\bibitem[{{Pinte} {et~al.}(2009){Pinte}, {Harries}, {Min}, {Watson},
  {Dullemond}, {Woitke}, {M{\'e}nard}, \& {Dur{\'a}n-Rojas}}]{Pinte2009}
{Pinte}, C., {Harries}, T.~J., {Min}, M., {et~al.} 2009, \aap, 498, 967

\bibitem[{{Pinte} {et~al.}(2006){Pinte}, {M{\'e}nard}, {Duch{\^e}ne}, \&
  {Bastien}}]{Pinte2006}
{Pinte}, C., {M{\'e}nard}, F., {Duch{\^e}ne}, G., \& {Bastien}, P. 2006, \aap,
  459, 797

\bibitem[{{Poglitsch} {et~al.}(2010){Poglitsch}, {Waelkens}, {Geis},
  {Feuchtgruber}, {Vandenbussche}, {Rodriguez}, {Krause}, {Renotte}, {van
  Hoof}, {Saraceno}, {Cepa}, {Kerschbaum}, {Agn{\`e}se}, {Ali}, {Altieri},
  {Andreani}, {Augueres}, {Balog}, {Barl}, {Bauer}, {Belbachir}, {Benedettini},
  {Billot}, {Boulade}, {Bischof}, {Blommaert}, {Callut}, {Cara}, {Cerulli},
  {Cesarsky}, {Contursi}, {Creten}, {De Meester}, {Doublier}, {Doumayrou},
  {Duband}, {Exter}, {Genzel}, {Gillis}, {Gr{\"o}zinger}, {Henning},
  {Herreros}, {Huygen}, {Inguscio}, {Jakob}, {Jamar}, {Jean}, {de Jong},
  {Katterloher}, {Kiss}, {Klaas}, {Lemke}, {Lutz}, {Madden}, {Marquet},
  {Martignac}, {Mazy}, {Merken}, {Montfort}, {Morbidelli}, {M{\"u}ller},
  {Nielbock}, {Okumura}, {Orfei}, {Ottensamer}, {Pezzuto}, {Popesso},
  {Putzeys}, {Regibo}, {Reveret}, {Royer}, {Sauvage}, {Schreiber}, {Stegmaier},
  {Schmitt}, {Schubert}, {Sturm}, {Thiel}, {Tofani}, {Vavrek}, {Wetzstein},
  {Wieprecht}, \& {Wiezorrek}}]{Poglitsch2010}
{Poglitsch}, A., {Waelkens}, C., {Geis}, N., {et~al.} 2010, \aap, 518, L2

\bibitem[{{Pontoppidan} \& {Dullemond}(2005)}]{Pontoppidan2005}
{Pontoppidan}, K.~M. \& {Dullemond}, C.~P. 2005, \aap, 435, 595

\bibitem[{{Ratzka} {et~al.}(2005){Ratzka}, {K{\"o}hler}, \&
  {Leinert}}]{Ratzka2005}
{Ratzka}, T., {K{\"o}hler}, R., \& {Leinert}, C. 2005, \aap, 437, 611

\bibitem[{{Rebolo} {et~al.}(1995){Rebolo}, {Zapatero Osorio}, \&
  {Mart{\'{\i}}n}}]{Rebolo1995}
{Rebolo}, R., {Zapatero Osorio}, M.~R., \& {Mart{\'{\i}}n}, E.~L. 1995, \nat,
  377, 129

\bibitem[{{Reipurth} \& {Clarke}(2001)}]{Reipurth2001}
{Reipurth}, B. \& {Clarke}, C. 2001, \aj, 122, 432

\bibitem[{{Riaz} \& {Gizis}(2012)}]{Riaz2012}
{Riaz}, B. \& {Gizis}, J.~E. 2012, \aap, 548, A54

\bibitem[{{Ricci} {et~al.}(2012){Ricci}, {Testi}, {Natta}, {Scholz}, \& {de
  Gregorio-Monsalvo}}]{Ricci2012}
{Ricci}, L., {Testi}, L., {Natta}, A., {Scholz}, A., \& {de Gregorio-Monsalvo},
  I. 2012, \apjl, 761, L20

\bibitem[{{Rodr{\'{\i}}guez-Ledesma} {et~al.}(2009){Rodr{\'{\i}}guez-Ledesma},
  {Mundt}, \& {Eisl{\"o}ffel}}]{Rodriguez2009}
{Rodr{\'{\i}}guez-Ledesma}, M.~V., {Mundt}, R., \& {Eisl{\"o}ffel}, J. 2009,
  \aap, 502, 883

\bibitem[{{Scholz} {et~al.}(2006){Scholz}, {Jayawardhana}, \&
  {Wood}}]{Scholz2006}
{Scholz}, A., {Jayawardhana}, R., \& {Wood}, K. 2006, \apj, 645, 1498

\bibitem[{{Spezzi} {et~al.}(2013){Spezzi}, {Cox}, {Prusti}, {Mer{\'{\i}}n},
  {Ribas}, {Alves de Oliveira}, {Winston}, {K{\'o}sp{\'a}l}, {Royer}, {Vavrek},
  {Andr{\'e}}, {Pilbratt}, {Testi}, {Bressert}, {Ricci}, {Men'shchikov}, \&
  {K{\"o}nyves}}]{Spezzi2013}
{Spezzi}, L., {Cox}, N.~L.~J., {Prusti}, T., {et~al.} 2013, \aap, 555, A71

\bibitem[{{Stamatellos} \& {Whitworth}(2009)}]{Stamatellos2009}
{Stamatellos}, D. \& {Whitworth}, A.~P. 2009, \mnras, 392, 413

\bibitem[{{Sz{\H u}cs} {et~al.}(2010){Sz{\H u}cs}, {Apai}, {Pascucci}, \&
  {Dullemond}}]{Szucs2010}
{Sz{\H u}cs}, L., {Apai}, D., {Pascucci}, I., \& {Dullemond}, C.~P. 2010, \apj,
  720, 1668

\bibitem[{{Testi} {et~al.}(2002){Testi}, {Natta}, {Oliva}, {D'Antona},
  {Comeron}, {Baffa}, {Comoretto}, \& {Gennari}}]{Testi2002}
{Testi}, L., {Natta}, A., {Oliva}, E., {et~al.} 2002, \apjl, 571, L155

\bibitem[{{Walker} {et~al.}(2004){Walker}, {Wood}, {Lada}, {Robitaille},
  {Bjorkman}, \& {Whitney}}]{Walker2004}
{Walker}, C., {Wood}, K., {Lada}, C.~J., {et~al.} 2004, \mnras, 351, 607

\bibitem[{{Weingartner} \& {Draine}(2001)}]{Weingartner2001}
{Weingartner}, J.~C. \& {Draine}, B.~T. 2001, \apj, 548, 296

\bibitem[{{Whelan} {et~al.}(2009){Whelan}, {Ray}, {Podio}, {Bacciotti}, \&
  {Randich}}]{Whelan2009}
{Whelan}, E.~T., {Ray}, T.~P., {Podio}, L., {Bacciotti}, F., \& {Randich}, S.
  2009, \apj, 706, 1054

\bibitem[{{Wilking} {et~al.}(2001){Wilking}, {Bontemps}, {Schuler}, {Greene},
  \& {Andr{\'e}}}]{Wilking2001}
{Wilking}, B.~A., {Bontemps}, S., {Schuler}, R.~E., {Greene}, T.~P., \&
  {Andr{\'e}}, P. 2001, \apj, 551, 357

\bibitem[{{Wilking} {et~al.}(1999){Wilking}, {Greene}, \&
  {Meyer}}]{Wilking1999}
{Wilking}, B.~A., {Greene}, T.~P., \& {Meyer}, M.~R. 1999, \aj, 117, 469

\bibitem[{{Wilking} {et~al.}(2005){Wilking}, {Meyer}, {Robinson}, \&
  {Greene}}]{Wilking2005}
{Wilking}, B.~A., {Meyer}, M.~R., {Robinson}, J.~G., \& {Greene}, T.~P. 2005,
  \aj, 130, 1733

\bibitem[{{Williams} \& {Cieza}(2011)}]{Williams2011}
{Williams}, J.~P. \& {Cieza}, L.~A. 2011, \araa, 49, 67

\bibitem[{{Young} {et~al.}(1986){Young}, {Lada}, \& {Wilking}}]{Young1986}
{Young}, E.~T., {Lada}, C.~J., \& {Wilking}, B.~A. 1986, \apjl, 304, L45

\end{thebibliography}

\Online
\begin{appendix}
\section{1$-$$\sigma$ upper limits for brown dwarfs not detected with \emph{Herschel}/PACS.}
\begin{table*}
\caption{1$-$$\sigma$ upper limits for brown dwarfs in $\rho$~Ophiuchi not detected in \emph{Herschel}/PACS images.}  
\label{table2}   
\centering          
\begin{tabular}{l l l l l l l l l}    
\hline\hline       
Identifier & RA & Dec & SpT  &  A$_{\emph{V}}$ & Ref. & 70$\mu$m & 100$\mu$m & 160$\mu$m \\ 
  	      &       &         &          &    (mag)         	 & 						& (mJy) & (mJy) & (mJy)   \\ 
\hline                    
  \multicolumn{9}{c}{Class~II} \\
\hline                  
GY92~15	 	       & 16:26:22.98 & $-$24:28:46.1 & M6    		& 10.7 & 7	  	& $<$47.5	& $<$324.9	& $<$579.8 \\	
GY92~109	        	& 16:26:42.89 & $-$24:22:59.1 & M6    		& 13.5 & 7	  	& $<$10.5	& $<$91.5	& $<$266.1 \\	
WL21	       			& 16:26:57.33 & $-$24:35:38.7 & M6    		& 23.8 & 7	  	& $<$120.9	& $<$223.1	& $<$631.4 \\	
GY92~350	        	& 16:27:46.29 & $-$24:31:41.2 & M6    		& 7.0  & 5	  	& $<$10.8	& $<$16.8	& $<$121.0 \\	
CFHTWIR-Oph~107 	& 16:28:48.71 & $-$24:26:31.8 & M6.25 	& 2.3  & 10 		& $<$4.4	& $<$10.0	& $<$46.4 \\	  
CFHTWIR-Oph~106 	& 16:28:29.93 & $-$24:54:06.4 & M6.5  		& 4.9  & 6	     	& $<$4.6	& $<$13.1	& $<$73.0 \\	  
CRBR~2322.3-1143   & 16:26:23.78 & $-$24:18:31.4 & M6.7  		& 8.6  & 4	     	& $<$11.2	& $<$38.9	& $<$202.5 \\	  
CFHTWIR-Oph~16 	& 16:26:18.58 & $-$24:29:51.6 & M6.75 	& 18.8 & 10 		& $<$30.3	& $<$50.4	& $<$227.5 \\	 
GY92~202	       		& 16:27:05.98 & $-$24:28:36.3 & M7(M6.5)  & 13.0 & 2,3  	& $<$39.7	& $<$50.2	& $<$227.5 \\	
CRBR~2317.3-1925   & 16:26:18.82 & $-$24:26:10.5 & M7.5(M5.5,M7)  	& 10.0 & 2,3,5  	& $<$89.3 & $<$119.9 & $<$449.7 \\	  
CFHTWIR-Oph~78   	& 16:27:26.23 & $-$24:19:23.1 & M7.75 	& 16.4 & 10 		& $<$8.1 	& $<$28.1	& $<$117.4 \\	  
CFHTWIR-Oph~96\tablefootmark{a}   	& 16:27:40.84 & $-$24:29:00.8 & M7.75 	& 2.4  & 10	& $<$23.5	& $<$113.6 & $<$396.3 \\	
GY92~64 	       		& 16:26:32.53 & $-$24:26:35.4 & M8    		& 11.0 & 2	     	& $<$35.5 	& $<$379.7	& $<$2414.2 \\	 
CFHTWIR-Oph~34   	& 16:26:39.92 & $-$24:22:33.6 & M8.25 	& 9.7  & 6	     	& $<$78.5 	& $<$63.5	& $<$270.2 \\	
GY92~11 	       		& 16:26:22.27 & $-$24:24:07.1 & M8.5(M6.5,M8.5)  	& 8.0  & 8,2,5 		& $<$10.4 & $<$50.1 & $<$361.5 \\	  
GY92~141	       		& 16:26:51.28 & $-$24:32:42.0 & M8.5(M8)  & 0.7  & 1,4   	& $<$8.6 	& $<$36.2 	& $<$133.0 \\	
CFHTWIR-Oph~98	& 16:27:44.20 & $-$23:58:52.4 & M9.75 	& 3.0  & 10 		& $<$5.9 	& $<$13.0 	& $<$46.3 \\	  
CFHTWIR-Oph~90	& 16:27:36.59 & $-$24:51:36.1 & L0    		& 2.4  & 10 		& $<$4.5 	& $<$12.8 	& $<$48.7 \\	  
\hline                  
  \multicolumn{9}{c}{Class~III} \\
\hline                  
GY92~450	        	& 16:28:03.56 & $-$24:34:38.6 & M6    		& 20.5 & 7  	& $<$4.4 	& $<$14.7 & $<$93.6 \\	  
CFHTWIR-Oph~4   	& 16:25:32.41 & $-$24:34:05.2 & M6.5  		& 2.5  & 6	& $<$10.9 	& $<$29.8 & $<$67.2 \\	 
CFHTWIR-Oph~101 	& 16:27:47.25 & $-$24:46:45.9 & M7    		& 25.8 & 10 	& $<$4.9 	& $<$11.9 & $<$102.9 \\	  
CFHTWIR-Oph~57	& 16:27:04.02 & $-$24:02:46.9 & M7.25 	& 6.1  & 6	& $<$7.8 	& $<$18.4 & $<$66.8 \\	 
CFHTWIR-Oph~47   	& 16:26:52.26 & $-$24:01:46.8 & M7.5  		& 5.6  & 6	& $<$6.8 	& $<$19.8 & $<$68.4 \\	 
CFHTWIR-Oph~77     & 16:27:25.64 & $-$24:37:28.6 & M9.75 	& 10.0 & 10 	& $<$6.1 	& $<$22.1 & $<$148.7 \\	 
CFHTWIR-Oph~103   & 16:28:10.46 & $-$24:24:20.4 & L0    		& 7.5  & 10 	& $<$5.5 	& $<$13.8 & $<$116.1 \\	 
\hline                  
  \multicolumn{9}{c}{No SED classification available} \\
\hline                  
SONYC-RhoOph-6        	& 16:27:05.93 & $-$24:18:40.2 & M7    		& 8.0  & 11   & $<$8.5 	& $<$28.4 & $<$133.8 \\	 
SONYC-RhoOph-1        	& 16:26:56.33 & $-$24:42:37.8 & M9    		& 5.0  & 9     & $<$5.5 	& $<$16.5 & $<$73.1 \\	  
CFHTWIR-Oph~9          	& 16:26:03.28 & $-$24:30:25.9 & L0    		& 6.6  & 10 	& $<$36.9 	& $<$28.5 & $<$476.0 \\	 
CFHTWIR-Oph~18         	& 16:26:19.41 & $-$24:27:43.9 & L0    		& 7.8  & 10 	& $<$11.7 	& $<$56.1 & $<$398.9 \\	         
CFHTWIR-Oph~100        	& 16:27:46.54 & $-$24:05:59.2 & L0    		& 8.0  & 10 	& $<$5.5 	& $<$15.2 & $<$58.5 \\	  
CFHTWIR-Oph~33         	& 16:26:39.69 & $-$24:22:06.2 & L4    		& 3.6  & 10 	& $<$165.2 	& $<$34.4 & $<$285.3 \\ 
\hline                  
  \multicolumn{9}{c}{Pre-brown dwarf} \\
\hline                  
Oph B-11        	& 16:27:13.96 & $-$24:28:29.3 &  	& $\sim$30 & 12   & $<$15.4 	& $<$32.0 & $<$63.1 \\	 
\hline                  
   \end{tabular}   
\tablefoot{\tablefoottext{a}{\citet{Harvey2012a} measured from their \emph{Herschel}/PACS ``mini-scan-map'' of this object F$_{\nu}$ (70$\mu$m)~$=$~5$\pm$20~mJy and F$_{\nu}$ (160$\mu$m)~$=$~130$\pm$200~mJy. These values are consistent with our upper limits.}}
\tablebib{(1) \citet{Luhman1997}; (2) \citet{Wilking1999}; (3) \citet{Luhman1999}; (4) \citet{Cushing2000}; (5) \citet{Natta2002}; (6) \citet{AlvesdeOliveira2010}; (7) \citet{McClure2010}; (8) \citet{Comeron2010}; (9) \citet{Geers2011}; (10) \citet{AlvesdeOliveira2012}; (11) \citet{Muzic2012}; (12) \citet{Andre2012}.}
   \end{table*}
\end{appendix}

\begin{appendix}
\section{Near-IR colour composite of GY92~344.}
\label{shadow}
\begin{figure*}
\centering
\includegraphics[width=\columnwidth]{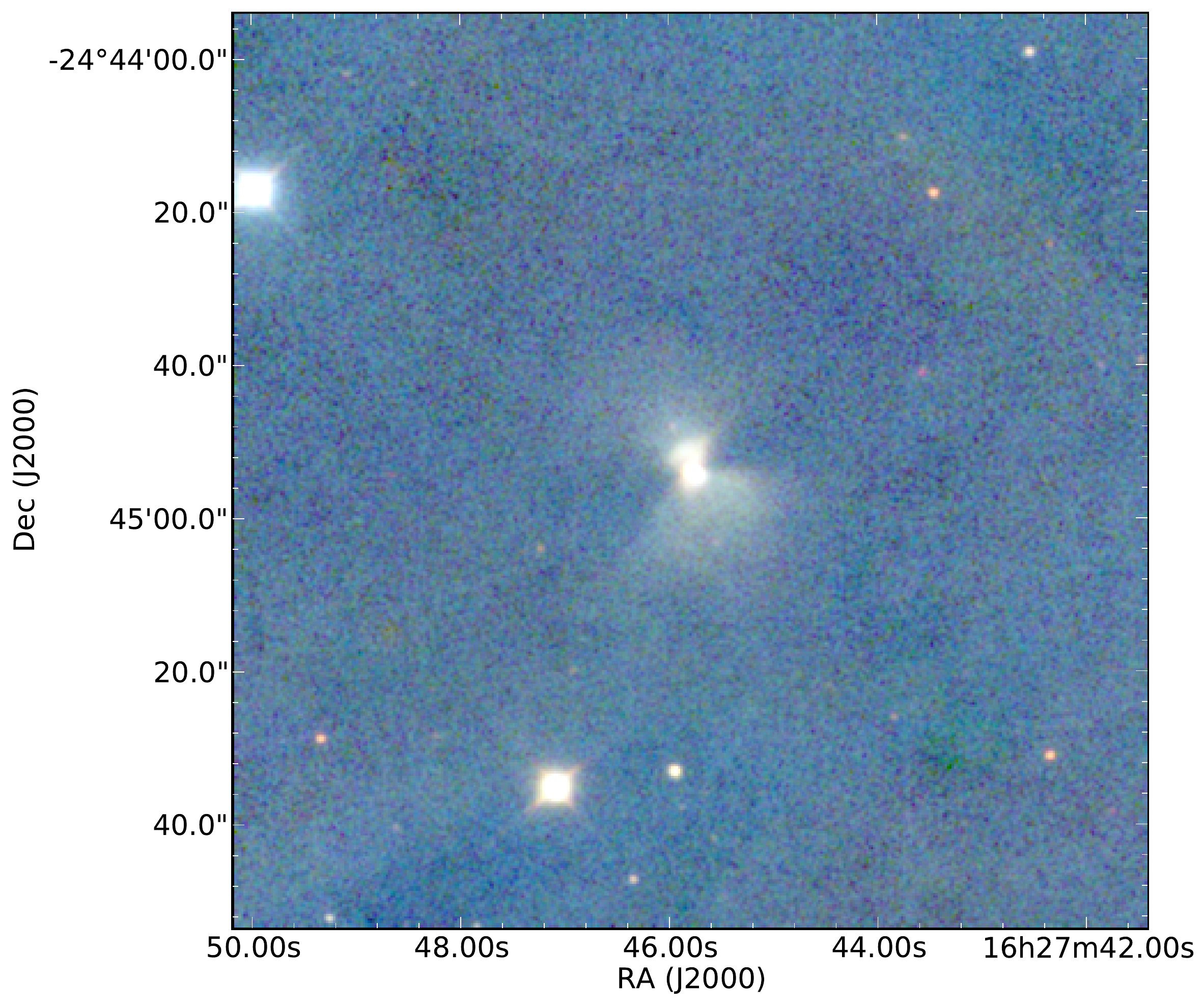}
\caption{WIRCam/CFHT \emph{JHK} colour-composite of GY92~344.}
\end{figure*}
\end{appendix}

\end{document}